\documentclass[%
groupedaddress,
preprint,
 amsmath,amssymb,
 aps,
prl,
]{revtex4-1}

\usepackage{graphicx}
\usepackage{dcolumn}
\usepackage{bm}
\usepackage{color}
\usepackage[normalem]{ulem}

\begin{document}


\title{Diffusive dynamics of contact formation 
in disordered polypeptides}

\author{G\"ul Zerze}
\affiliation
{Department of Chemical and Biomolecular Engineering, 
Lehigh University, Bethlehem PA}
\author{Jeetain Mittal}
\email{jeetain@lehigh.edu}
\affiliation
{Department of Chemical and Biomolecular Engineering, 
Lehigh University, Bethlehem PA}
\author{Robert B. Best}
\email{robertbe@helix.nih.gov}
\affiliation
{Laboratory of Chemical Physics, National Institute of Diabetes and
Digestive and Kidney Diseases, National Institutes of Health,
Bethesda MD 20892}

%
%
%

\date{\today}

\begin{abstract}
Experiments measuring contact formation between a probe and quencher
in disordered chains provide
information on the fundamental dynamical timescales relevant to 
protein folding, but their interpretation usually relies on simplified
one-dimensional (1D) diffusion models. Here, we use all-atom molecular
simulations to capture both the time-scales of contact formation, as well as
the scaling with the length of the peptide for tryptophan triplet
quenching experiments. Capturing the experimental quenching times depends
on the water viscosity, but more importantly on the configurational space explored by 
the chain. We also show that very similar results are obtained from 
Szabo-Schulten-Schulten theory applied to a 1D diffusion model 
derived from the simulations, supporting the validity of 
such models. However, we also find a significant reduction in 
diffusivity at small separations, those which are most
important in determining the quenching rate.

\begin{description}
\item[Usage]
Secondary publications and information retrieval purposes.
\item[PACS numbers]
May be entered using the \verb+\pacs{#1}+ command.
\item[Structure]
You may use the \texttt{description} environment to structure your abstract;
use the optional argument of the \verb+\item+ command to give the category of each item. 
\end{description}
\end{abstract}

\pacs{Valid PACS appear here}
\maketitle
    
%

Characterizing the configuration distribution and dynamics within unfolded
or disordered peptides is a first step toward understanding more complex processes
such as protein folding and aggregation~\cite{kubelka-2004}. 
To this end, contact quenching pump-probe experiments are a sensitive measure
of dynamics in disordered peptides,
which can be used to determine
loop formation rates~\cite{bieri-1999,lapidus-2000,lapidus-2002,buscaglia-2006}, 
helix-coil dynamics~\cite{lapidus-2002-2,fierz-2009} and even the folding rate of
small proteins~\cite{buscaglia-2005}. 
In these experiments, a probe is initially excited to a long-lived electronic
state which can be quenched by contact with a second species distant in 
sequence, allowing the chain dynamics to be monitored.
However, interpretation of the data
usually requires fairly strong assumptions about the nature of the probe-quencher
distance distribution and dynamics; inclusion of additional experimental data 
such as F\"orster resonance energy transfer (FRET) efficiencies can help to 
constrain the distance distributions~\cite{soranno-2009}.

An alternative approach to interpreting contact formation experiments is to 
use molecular simulations to compute 
quenching rates directly \cite{yeh-2002}, requiring only the
knowledge of the distance dependence of the contact quenching rate.
In principle, these can provide a detailed view of the chain dynamics, without
the need for simplifying assumptions. 
Previous insightful work 
using atomistic simulations has been used to interpret contact
quenching rates in short disordered peptides. It was found that the rates
obtained from simulation needed to be reduced by a factor of 2-3
in order to match experiment, which was attributed to the viscosity 
of the water model being too low in the simulations~\cite{yeh-2002}. 
However, this also
assumes that all rates, including the quenching rate, are slowed
by the same viscosity factor, which may not be realistic.
In addition, the interpretation is complicated by the
over-collapsed nature of the disordered ensemble,
relative to estimates from experimental measurements
such as FRET, SAXS or light scattering 
for many atomistic force fields
~\cite{nettels-2009,piana-2014,best-2014-2}.
Recent physically motivated refinements of protein force fields,
have been shown to yield more accurate equilibrium properties for disordered
chains~\cite{best-2014-2,piana-2015}, and should not require any correction for
viscosity, owing to the use of more accurate water models 
~\cite{abascal-2005,piana-2015}.
These should alleviate the need for adjustments to the data, and allow a more
direct interpretation of the experimental results.

\begin{figure}[b]
\centering
\includegraphics[width=2.5in]{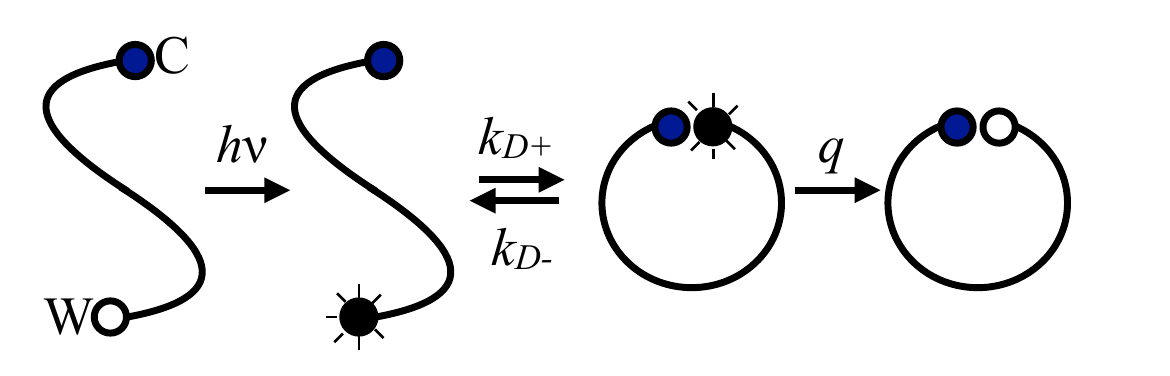}
\caption{Kinetic scheme for triplet quenching: After 
initial laser excitation of the tryptophan to the triplet state,
the terminal residues may come into contact
with a diffusion-controlled rate $k_{D+}$. The
triplet is quenched in contact with rate $q$. The termini may
also separate with rate $k_{D-}$.}
\label{fig:schematic}
\end{figure}

Here, we focus on a set of experiments in which the dynamics of a series of 
peptides of composition C(AGQ)$_n$W-NH$_2$ (hereafter: AGQ$_n$)
was monitored from the rate at which the triplet state of the 
tryptophan (W) at one end of the chain was quenched by van der Waals contact with the 
cysteine residue (C) at the other end \cite{lapidus-2000,lapidus-2002,buscaglia-2006}.
The experiment is illustrated by the 
kinetic scheme in Fig. \ref{fig:schematic}. Briefly, after optical
excitation, the termini of the peptide will diffuse relative
to each other, and may be quenched on contact. In the extreme
``diffusion-limited'' scenario, the quenching on contact is
so fast that the observed rate of triplet quenching $k_\mathrm{obs}$ 
is just the 
diffusion-limited rate of contact formation, $k_{D+}$. In 
the opposite ``reaction-limited'' extreme, 
quenching is very slow and the termini must
contact many times on average before a quenching event occurs,
in which case the overall quenching rate depends only on
the population of the contact states. The actual rate of 
quenching is usually somewhere between these scenarios,
and so contains information on both the distance distribution
and the dynamics of quenching.

We have carried out extensive molecular dynamics simulations of a
series of 
AGQ$_n$ peptides, for $n=1-6$, using three related
force fields: the Amber ff03* protein force field \cite{duan-2003,best-2009-2} 
together with the 
TIP3P water model\cite{jorgensen-1983}; the Amber ff03w protein force field
\cite{best-2010-3}, which is used in combination with a
more accurate water model, TIP4P/2005 
\cite{abascal-2005}; and the Amber ff03ws protein force field
\cite{best-2014-2} which also uses TIP4P/2005 water, but with 
strengthened protein water interactions. 
Specifically, in ff03ws
the values of Lennard-Jones $\epsilon$ for all protein-water atom pairs
are scaled by a factor 1.10 relative to the standard combination rule,
in order to correct the 
overly-collapsed nature of the disordered ensemble
\cite{best-2014-2}. 
The simulations were 
run using Gromacs 4.5 or 4.6 \cite{hess-2008} 
at a constant pressure of 1 bar 
and a constant temperature of 293 K for a total time of
2-10 $\mu$s for each peptide and force field.
Initial
conditions were obtained either from short temperature replica
exchange simulations, or from high-temperature runs at constant
volume (see electronic supplementary information (ESI) for full 
details).

\begin{figure}[htbp]
\centering
\includegraphics[width=3.25in]{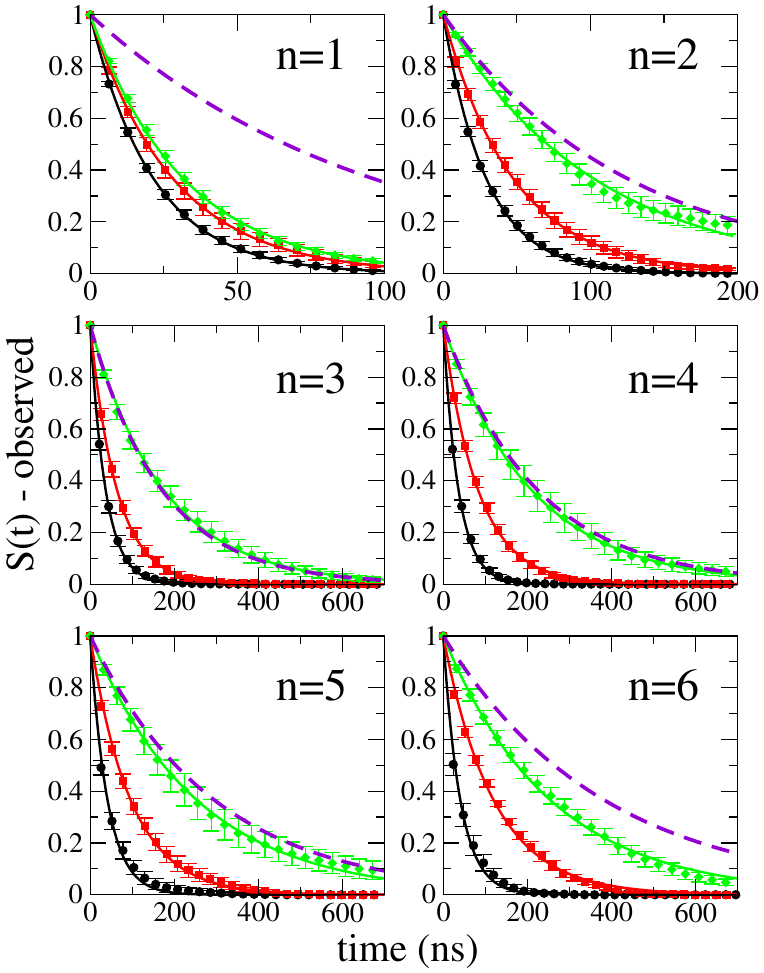}
\caption{Decays of tryptophan triplet state. Overall decays 
calculated from simulations with the ff03* (black), ff03w (red),
and ff03ws (green) force fields are shown for peptides AGQ$_n$
for $n=1-6$, together with the corresponding experimental decays
(broken purple lines). Symbols: simulation data; lines: single
exponential fits to data.}
\label{fig:st}
\end{figure}

In order to compare our results directly with experiment,
we compute quenching rates
using a step function for the dependence of the
quenching rate $q$ on the Trp-Cys separation
$r_\mathrm{cw}$, that is 
$q(r_\mathrm{cw}) = q_c H(r_c-r_\mathrm{cw})$, 
where $q_c = 8\times10^{8}$ s$^{-1}$ 
is the constant quenching
rate in contact, $H(x)$ is the Heaviside step function and $r_c=0.4$ nm
is the contact distance. The distance $r_\mathrm{cw}$ is taken as the minimum 
distance between the sulfur in the cysteine side-chain and the
heavy atoms of the tryptophan indole ring system~\cite{lapidus-2000,yeh-2002}. 
The observed quenching rate is then determined from 
the decay of the triplet survival probability 
$S(t) = \langle \exp[-\int_{t_0}^{t_0+t} q(r_\mathrm{cw}(t')) dt']
\rangle_{t_0}$, where 
the average is over equilibrium initial conditions $t_0$, obtained
by taking every saved frame of the 
simulation as a valid starting point. 

Overall decay curves for the triplet population
determined using the step function 
form for $q(r_\mathrm{cw})$ are shown in Figure \ref{fig:st}, compared with 
experimental decays \cite{lapidus-2002}. The simulation data for Amber ff03ws 
is in excellent agreement with the experiment for all $n$ values except $n=1$. 
As expected, there are
large differences amongst the force fields, with quenching rates
for ff03ws being
significantly slower than for those ff03* and ff03w. Part of the difference
between ff03* and ff03ws is expected to be the $\sim3$-fold lower viscosity
of the TIP3P water model relative to
TIP4P/2005 (the latter being very close to the true value)
\cite{yeh-2002}. 
However, this is clearly not the only effect, since the decay
for the ff03w force field, which also uses TIP4P/2005 water,
is only slightly slower than that for ff03*. Therefore, the
change in equilibrium conformational distribution from ff03w to ff03ws
must also play a role in the observed difference.

\begin{figure}[htbp]
\centering
\includegraphics[width=3.25in]{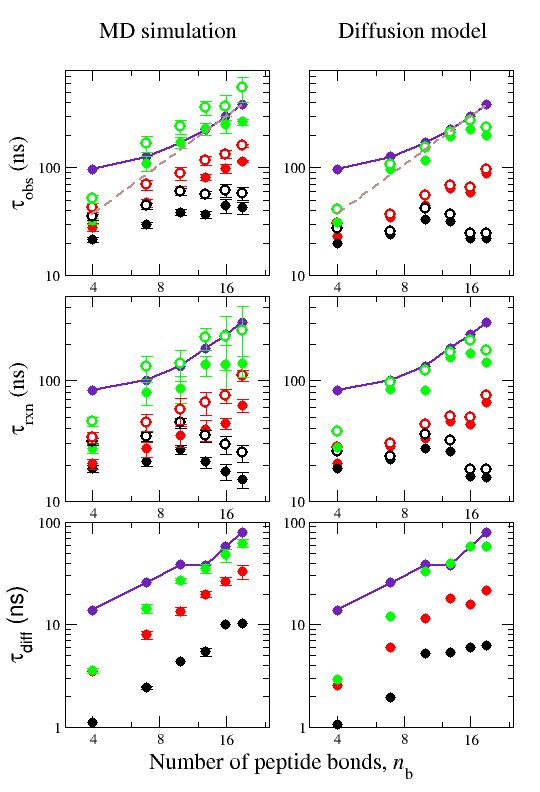}
\caption{Dependence of quenching times (inverse of quenching rates)
on chain length. Top, middle and bottom rows show the 
overall, reaction-limited and diffusion-limited quenching times, respectively.
Simulation data for Amber ff03*, ff03w and ff03ws are shown 
by black, red and green symbols, respectively. Filled and empty symbols are
for step-function and exponential distance dependence respectively. 
Experimental data are shown by 
purple symbols, and $n_{\mathrm{b}}^{3/2}$ scaling expected for a Gaussian
chain by the broken line. Left panels are rates calculated
directly from simulation and right panels are those calculated
from 1D diffusion model using SSS theory. 
}
\label{fig:ratefig}
\end{figure}

We summarize the peptide-length dependence of the observed
quenching rate in Figure \ref{fig:ratefig}. This confirms that 
the observed rate is in excellent agreement with the experimental
data for ff03ws, 
while at the opposite extreme, ff03* results
in rates which are almost independent of peptide length. 
The
ff03ws results approximately follow an $n_\mathrm{b}^{3/2}$ dependence
of the reaction-limited quenching time on the number of
peptide bonds $n_\mathrm{b}$, as expected for a
Gaussian chain \cite{lapidus-2002}. 
The power law which best fits the data is $n_\mathrm{b}^{1.38\pm0.12}$ 
which is also in agreement with the trend in experiment toward
$n_\mathrm{b}^{3/2}$ for longer
AGQ sequences~\cite{lapidus-2002}, as well with the fit
to data for a different peptide sequence 
($n_\mathrm{b}^{1.36\pm0.26}$)~\cite{bieri-1999}. Interestingly, these quenching
rates exhibit a very different scaling compared with loop formation rates
in single stranded DNA \cite{cheng-2010}.


We can obtain more insight into the contributions to the 
observed relaxation rate by splitting it into
diffusion-controlled and reaction-controlled parts
\cite{lapidus-2002}, via
$k_\mathrm{obs}^{-1} = k_\mathrm{D+}^{-1} + k_\mathrm{R}^{-1}$.
We determine the reaction-limited $k_R$ by integrating
over the Trp-Cys distance distribution, 
$k_\mathrm{R}=\int_0^\infty q(r_\mathrm{cw}) P(r_\mathrm{cw}) dr_\mathrm{cw}$. 
The diffusion
limited rate can be obtained by using a step function for
the survival probability $S(t)$ and averaging over time origins $t_0$,
$S(t) =\langle H(t_c(t_0)-t-t_0) \rangle_{t_0}$.
Here, $t_c(t_0)$ is the first time after $t_0$ when the Trp and
Cys contact, $H(x)$ is again the Heaviside step function, 
and all time points in the simulation where the probes
are not already in contact are used as separate time origins $t_0$.
This calculation (Figure \ref{fig:ratefig})
reveals that for all of the peptides, the 
observed rates are in fact much closer to the reaction-limited
rates, although there is a non-negligible contribution from 
diffusion. Interestingly, the slowdown in the diffusion-limited
rate from ff03* to ff03w (by changing from TIP3P to
TIP4P/2005) is very close to the 2-3 fold expected from the
change of water viscosity. There is an additional slowdown
in the diffusion limited rate when moving from ff03w to 
ff03ws, which presumably arises from the larger configurational
space which must be explored due to the more expanded chain~\cite{best-2014-2}. 
In summary, it is clear
that most of the improved agreement with experiment which
we obtain by using Amber ff03ws comes from the reaction-limited
rate. 

In the above analysis, we have assumed a very simple distance
dependence of the rate of quenching. 
As an alternative, assuming that the quenching occurs via an
electron transfer mechanism \cite{desancho-2011},
we use an exponential distance-dependent
rate $q(r_\mathrm{cw}) = k_0\exp[\beta(r_\mathrm{cw}-r_c)]$
where $r_c$ is the same 
and we have fitted the parameters
$k_0 = 1\times10^{8}$ s$^{-1}$ and $\beta = 33.33$ nm$^{-1}$
to bimolecular quenching data 
for tryptophan and cysteine embedded in a glass \cite{lapidus-2001} 
(see ESI for full details). 
The resulting rates, shown by empty symbols in Fig. \ref{fig:ratefig}, 
are slightly slower, but generally in good agreement with 
those obtained from the step function distance dependence,
suggesting that our conclusions are not overly sensitive to the
particular form used for the rate. 
Considering the extremely 
sharp distance-dependence of the quenching rate, it is
quite reasonable that the step function can be a good approximation. 

\begin{figure}
\centering
\includegraphics[width=3.25in]{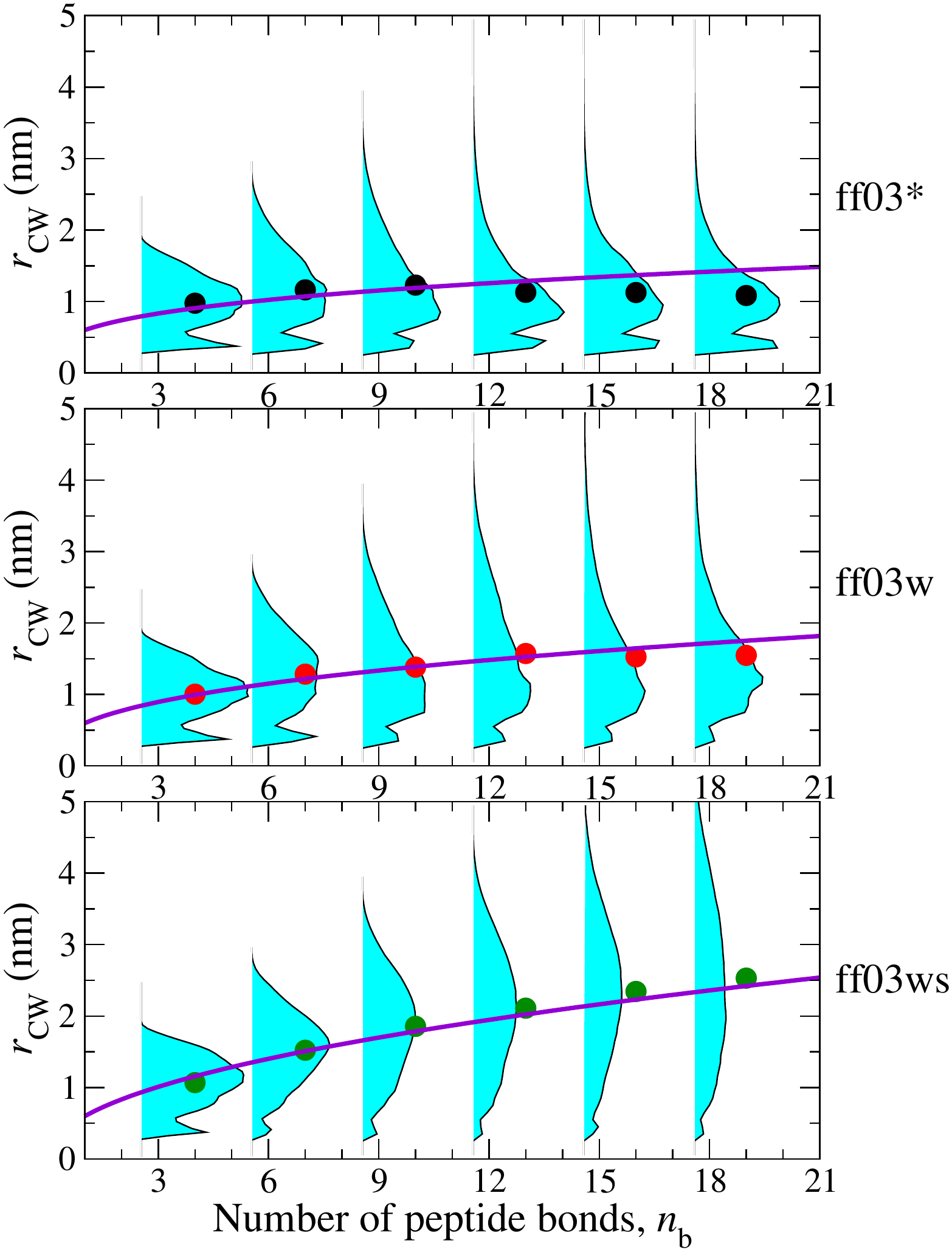}
\caption{Distribution of Trp-Cys distance for each chain 
length $n_b$ for AGQ$_n$ peptides, for three force fields. Symbols show
the mean distance and curves are power law fits.}
\label{fig:freeg}
\end{figure}

To understand the relationship between the chain dynamics and its 
structural properties, we characterize the equilibrium ensemble of conformations 
sampled in these simulations by the distribution of distances between
the tryptophan and cysteine, shown in Figure \ref{fig:freeg}. These reveal distinct differences
amongst the different force fields, which become more apparent
as the number of AGQ repeats is increased: the simulations 
with Amber ff03* and ff03w tend to be quite collapsed, while those
with ff03ws are relatively expanded.  
Notably, the mode of the distance distributions for ff03* hardly shifts 
as a function of chain length $n$, remaining near $\sim 1$ nm. For ff03w, weak 
expansion of the chain as a function of $n$ from $\sim 1$ to $1.5$ nm is observed.
In contrast, 
AGQ$_n$ expands with $n$ for the ff03ws force field as expected
for a chain in good solvent. 

We have quantified the polymer scaling properties
of AGQ$_n$ peptides by fitting the dependence of the mean Trp-Cys distance 
on the number of peptide bonds $n_\mathrm{b}$ to a power law 
$r_\mathrm{cw}=A n_\mathrm{b}^\nu$ (similar results are obtained
using the end-to-end distance), with $A$ fixed to 0.6 nm for
all peptides.
The exponents of 0.30 (0.02) and 0.36 (0.01) for ff03* and
ff03w respectively are indicative of a chain in poor solvent~\cite{mao-2010}, 
while the ff03ws exponent of 
0.47 (0.01) is 
close to the average exponent of 0.46 (0.05) determined experimentally 
for unfolded and disordered proteins~\cite{hofmann-2012}.
The trends for the reaction limited rates are consistent
with the equilibrium distance distributions, with the
collapsed ff03* and ff03w being very similar to one another,
and relatively independent of chain length. 
An important difference between these two force fields is that the
reaction-limited rates for ff03* even slightly increase with $n_\mathrm{b}$, 
which is not expected.
Lastly, we note that an important distinction relative to the 
distributions frequently assumed in interpreting 
experiments~\cite{lapidus-2002,buscaglia-2006}, is the existence of a separate short-range 
peak for the contact population in Fig.~\ref{fig:freeg}. 
The relative orientation of the Trp and Cys
appears to be broad with no strongly preferred interaction modes
(see ESI). The lifetime
of this population is 0.4-0.9 ns for the ff03ws force field,
depending on the peptide.

\begin{figure}
\centering
\includegraphics[width=8 cm]{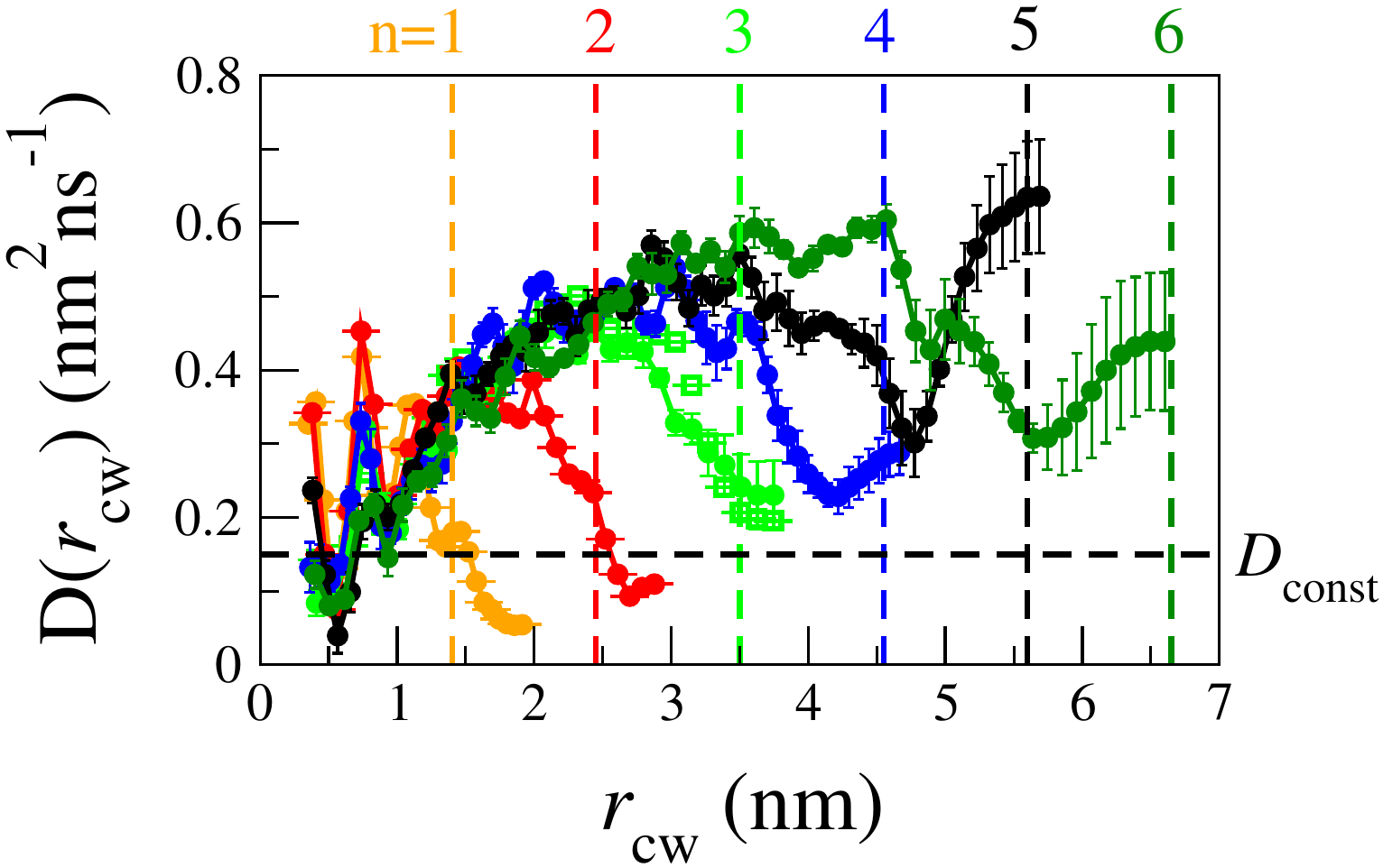}
\caption{Position-dependent diffusion coefficients $D(r_\mathrm{cw})$, 
for each peptide AGQ$_n$, $n=1-6$. Error bars calculated from division
of data into 5 non-overlapping blocks. Vertical lines indicate the 
backbone extension for a fully extended chain. Empty symbols for
$n=3$ are results from a 5 nm simulation box (vs 4 nm for solid
symbols). Horizontal line is constant diffusion
coefficient $D_\mathrm{const}$ needed to fit the data for $n=3-6$.}
\label{fig:sss}
\end{figure}

Next, we test an approximation commonly used to
analyze experimental data on contact formation,  
namely that 
the dynamics of the chain can be approximated as
one-dimensional diffusion along the Trp-Cys distance
coordinate. 
To investigate the accuracy
of such a model for capturing the dynamics in the full
phase space, we have fitted our simulation data 
$r_\mathrm{cw}(t)$ to a 1D diffusion model, using an established
Bayesian approach \cite{hummer-2005,best-2006-2,mittal2008layering,mittal2012pair}. 
Briefly, the method attempts to find the diffusive
model, defined by 
a potential of mean force 
$F(r_\mathrm{cw}) = -\ln p_\mathrm{eq}(r_\mathrm{cw})$ and position-dependent
diffusion coefficients $D(r_\mathrm{cw})$, whose propagators
best match the observed history of the simulations
(details in ESI). The diffusion coefficients thus obtained
are shown in Figure \ref{fig:sss} for ff03ws as a function of 
the number of AGQ repeats $n$ in the peptides. 

The diffusion coefficients we estimate are very 
comparable to those obtained for the same peptide
from direct analysis of contact 
quenching data, $\sim 0.2$ nm$^{2}$ns$^{-1}$\cite{lapidus-2002},
and from MD simulations, $0.3-0.9$ nm$^{2}$ns$^{-1}$\cite{yeh-2002}
(after considering the low viscosity of the TIP3P water model used), 
as well as with diffusion coefficients estimated for an 
unfolded protein from single
molecule FRET in water $\sim0.1$ nm$^{2}$ns$^{-1}$
\cite{nettels-2007}. However, our analysis reveals a significant
distance-dependence to the diffusion coefficient not included in
prior work. Specifically, the diffusion coefficients vary 
relatively little at large separations, but strongly 
decrease at short probe-quencher distances,
most likely due to the increased chain density at small
distances, as well as hydrodynamic effects as the Trp
and Cys approach each other. Remarkably, the $D(r_\mathrm{cw})$
curves are nearly superimposable for short and intermediate
separations $r_\mathrm{cw}$ of Trp and Cys, for all of
the peptides. Each peptide 
deviates from this common curve only when it 
approaches its maximum extension
(vertical broken lines in Fig. \ref{fig:sss}). 
This
is not a finite size effect, as we obtain almost identical
results for $n=3$ with a larger simulation box 
(Fig. \ref{fig:sss}). 

Although we have been able to determine a ``best-fit'' one-dimensional
model, this by itself does not guarantee that the dynamics of
this model is faithful to that of the full simulation (projected
onto the same coordinate). To check this, we have used
Szabo-Schulten-Schulten (SSS) theory\cite{szabo-1980} to compute rates from the
diffusion model for all simulations and compare them with those computed without
dynamical approximations from the simulations
(details in ESI).
The results, shown in Figure \ref{fig:ratefig} are in excellent agreement
with the direct analysis of the simulations. 
We can also try to estimate the effective constant diffusion coefficient 
$D_\mathrm{const}$
which would be required to match the measured 
diffusion-limited quenching rates. 
For $n=1-2$, we obtain
$D_\mathrm{const}\sim0.3$ nm$^2$~ns$^{-1}$, and for $n=3-6$,
$D_\mathrm{const}\sim0.15$ nm$^2$~ns$^{-1}$ (the latter value
in close agreement with the experimental estimate~\cite{lapidus-2002} of 
$\sim0.17$ nm$^2$~ns$^{-1}$). 
This is also expected from ESI eq. 2 as  
the $D(r)$ at short separations are effectively weighted much
more, thereby helping to rationalize the experimental observation that
the diffusion coefficient for describing contact formation
is about an order of magnitude smaller than the relative 
bimolecular Trp-Cys diffusion coefficient~\cite{lapidus-2002}.

Our results indicate that treating
contact formation using simple 1D diffusion models
captures the relevant dynamics accurately, justifying the
use of such models in interpreting experiment. This is a remarkable
result because there are many situations where the end-end
distance is in fact not a good reaction coordinate, except
in the presence of a mechanical pulling force \cite{dudko-2011,morrison-2011}. 
However, the simulation results do suggest additional complexity
beyond what could reasonably be assumed \textit{a priori} when interpreting
the experimental data: namely that 
the distance distribution functions $P(r_\mathrm{cw})$ include 
an additional peak at short separations corresponding to the 
contacting residues, 
and the diffusion coefficients $D(r_\mathrm{cw})$
exhibit a strong distance-dependence at the short separations, which
are most important for determining the diffusion-limited rate of 
contact formation: indeed the effective 
position-independent diffusion coefficients 
obtained by fitting SSS theory to experimental quenching rates would be
almost entirely determined by the diffusion coefficients at the shortest
probe-quencher separations. These results should aid in the interpretation
of future contact quenching experiments, as well as the many other
types of experiment monitoring a single residue-residue distance, 
which are often also modeled as using 1D diffusion: these include
single molecule FRET, optical tweezers, and
atomic force microscopy experiments. 




\noindent\textbf{Acknowledgements}
We thank Marco Buscaglia, Bill Eaton, Gerhard Hummer and Ben Schuler  
for helpful comments on the manuscript. We acknowledge support from the U.S. Department of Energy, Office of Basic Energy Science, Division of Material Sciences and Engineering under Award (DE-SC0013979) (J.M.) and the Intramural Research Program of the National Institute of Diabetes and Digestive and Kidney Diseases of the National Institutes of Health (R.B.B.). This study utilized the high-performance computational capabilities of the Biowulf Linux cluster at the National Institutes of Health, Bethesda, Md. (http://biowulf.nih.gov) and Extreme
Science and Engineering Discovery Environment (XSEDE), grant no. TG-MCB-120014.

\bibliography{all}

\end{document}


\preprint{APS/123-QED}

\title{Electronic Supporting Information for: 
Diffusive dynamics of contact formation in disordered polypeptides}

\author{G\"ul Zerze}
\affiliation
{Department of Chemical and Biomolecular Engineering, 
Lehigh University, Bethlehem PA 18034}
\author{Jeetain Mittal}
\email{jeetain@lehigh.edu}
\affiliation
{Department of Chemical and Biomolecular Engineering, 
Lehigh University, Bethlehem PA 18034}
\author{Robert B. Best}
\email{robertbe@helix.nih.gov}
\affiliation
{Laboratory of Chemical Physics, National Institute of Diabetes and
Digestive and Kidney Diseases, National Institutes of Health,
Bethesda MD 20892}

%
%
%

\date{\today}

\pacs{Valid PACS appear here}
\maketitle


\section{Details of molecular simulations}

In order to obtain initial conditions of serial simulations of n=1,2 peptides, 
we ran 50 ns/replica parallel tempering in well-tempered ensemble 
(PTWTE)~\cite{bonomi2010enhanced,deighan2012efficient} simulations for both n=1 and n=2 peptide, 
using 1.2 kJ/mol and 500 kJ/mol as the Gaussian height and width, respectively. 
Eight replicas were used in a temperature range from 300 K to 503 K, 
using a bias factor of 8. We keep the 300 K replica neutral, i.e. no energy 
bias is applied, while allowing it to exchange with the adjacent replica 
(a similar application as in~\cite{sutto2013effects}). The last 24 frames of the
300 K replica were picked, which were saved in every 2 ns, to serve as the initial 
conditions of serial runs.
\begin{table}[htbp]
\centering
\begin{tabular}{l c c c c c c }
\hline
\hline
        & \multicolumn{2}{c}{ff03*} &  \multicolumn{2}{c}{ff03w} & \multicolumn{2}{c}{ff03ws}  \\
Peptide & runs & time/$\mu$s & runs & time/$\mu$s & runs & time/$\mu$s \\
\hline
AGQ$_1$ & 24 & 2.4  & 24 & 2.4  & 24 & 2.4  \\
AGQ$_2$ & 24 & 4.8  & 24 & 4.8  & 24 & 4.8  \\
AGQ$_3$ & 8 & 7.9  & 8 & 7.6  & 8 & 7.6  \\
AGQ$_4$ & 8 & 7.8  & 8 & 7.5  & 8 & 7.4  \\
AGQ$_5$ & 8 & 7.7  & 8 & 7.5  & 8 & 7.5  \\
AGQ$_6$ & 5 & 4.4  & 5 & 4.2  & 10 & 9.6  \\
\hline
\end{tabular}
\caption{Number of runs and total length of runs for each force field
and peptide}
\label{tab:run_stats}
\end{table}

Initial conditions for peptides with $n=3-6$ were obtained 
by running a 200 ps constant pressure simulation at 293 K to equilibrate the
box size, followed by a 20 ns constant volume simulation at 500 K. The 
last frame of the 500 K run was used to initiate long runs at constant
pressure for computing the quenching rates.
Simulations were run in one of three force field combinations:
(i) Amber ff03*\cite{best-2009-2} with TIP3P water\cite{jorgensen-1983},
(ii) Amber ff03w\cite{best-2010-3} with TIP4P/2005 water 
\cite{abascal-2005}, and 
(iii) Amber ff0ws\cite{best-2014-2}, also with TIP4P/2005 water.
Molecular dynamics simulations were run at constant pressure of 1 bar using a 
Parinello-Rahman barostat with a coupling time of 5 ps and
a compressibility of $4.5\times10^{-5}$ bar$^{-1}$, and 
constant temperature of 293 K using a 
velocity-rescaling thermostat \cite{bussi-2007}
with coupling time of 1 ps. This thermostat was chosen because
it should have a minimal effect on the dynamics (compared to
e.g. Langevin dynamics at high friction). Lennard-Jones interactions were 
evaluated every step for atoms separated by up to 0.9 nm and
every 10 steps for atoms separated by between 0.9 and 1.4 nm.
Mean-field corrections to energy and pressure were included
for Lennard-Jones interactions beyond 1.4 nm. Long-range
electrostatics were included via the particle-mesh Ewald 
scheme \cite{darden-1993} with a grid spacing of $\sim 0.12$ nm 
and a real-space cut-off of 0.9 nm. 
A summary of the number of runs and cumulative simulation time
for each peptide and force field are listed in Table \ref{tab:run_stats}.

\section{Distribution of contact configurations}

\begin{figure}[htbp]
\centering
\includegraphics[width=8 cm]{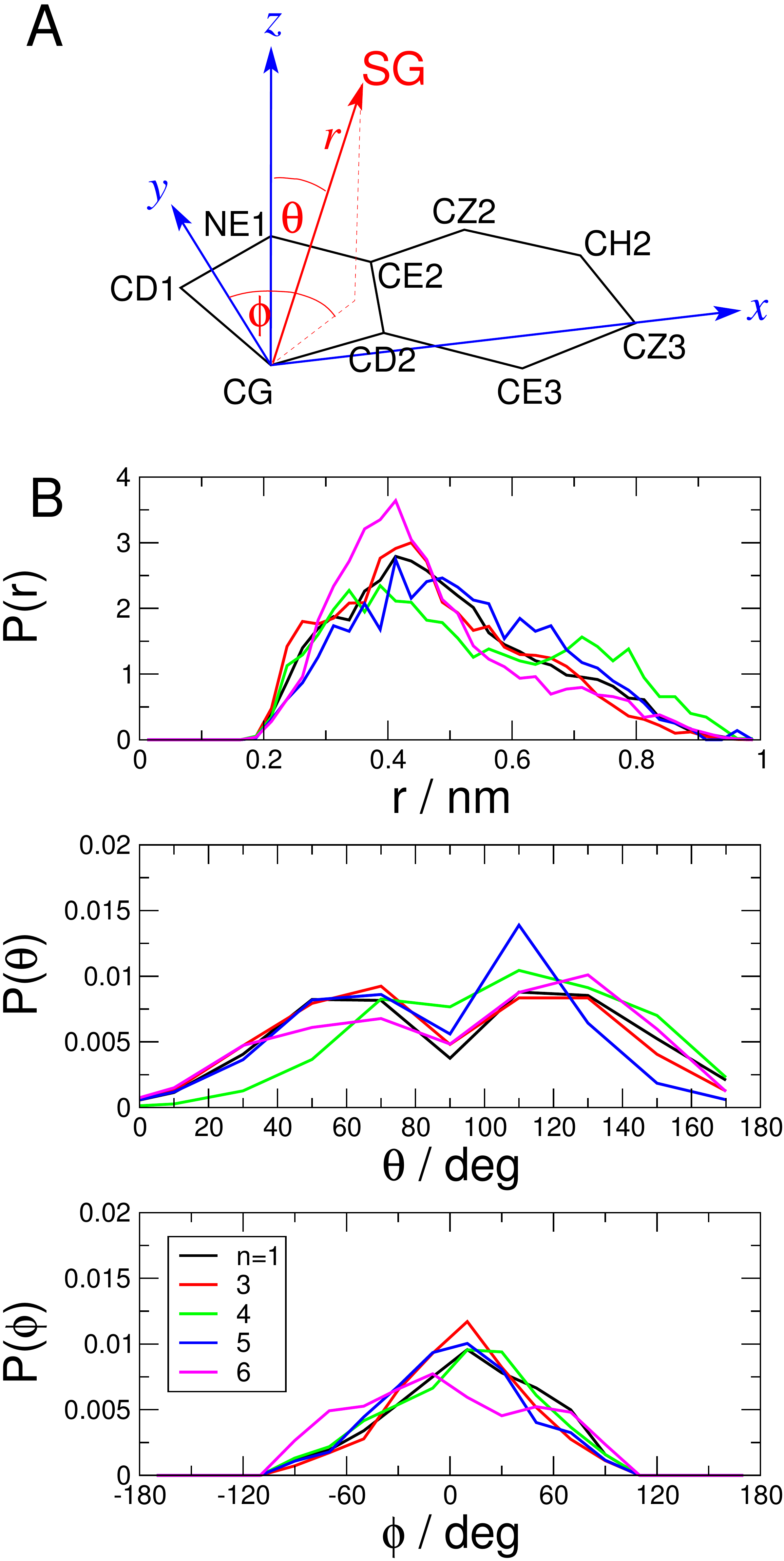}
\caption{Distribution of Trp-Cys contact configurations. (A) The 
position of the Cys SG is defined in a frame of reference
relative to the Trp. (B) Distributions of spherical polar coordinates
for SG in this axis frame.}
\label{fig:orient}
\end{figure}

To characterize the modes of association of Trp and Cys, we
determine the distribution of the position of the Cys sulfur (SG)
in a frame of reference defined by the Trp. This frame has
the origin at the Trp CG carbon and the $x$-axis running through
the CZ3 carbon. The $z$-axis is parallel to the vector
$x\times t$, where $t$ is the vector from the Trp CG to NE1,
and $y = z \times x$ (Fig. \ref{fig:orient} A). The corresponding
distributions of spherical polar coordinates suggest there is no
strongly preferred contact orientation (Fig. \ref{fig:orient} B),
and that Trp and Cys interact in similar orientations in all
peptides.

\section{Fitting of diffusion models using Bayesian method}

\begin{figure}[htbp]
\centering
\includegraphics[width=8 cm]{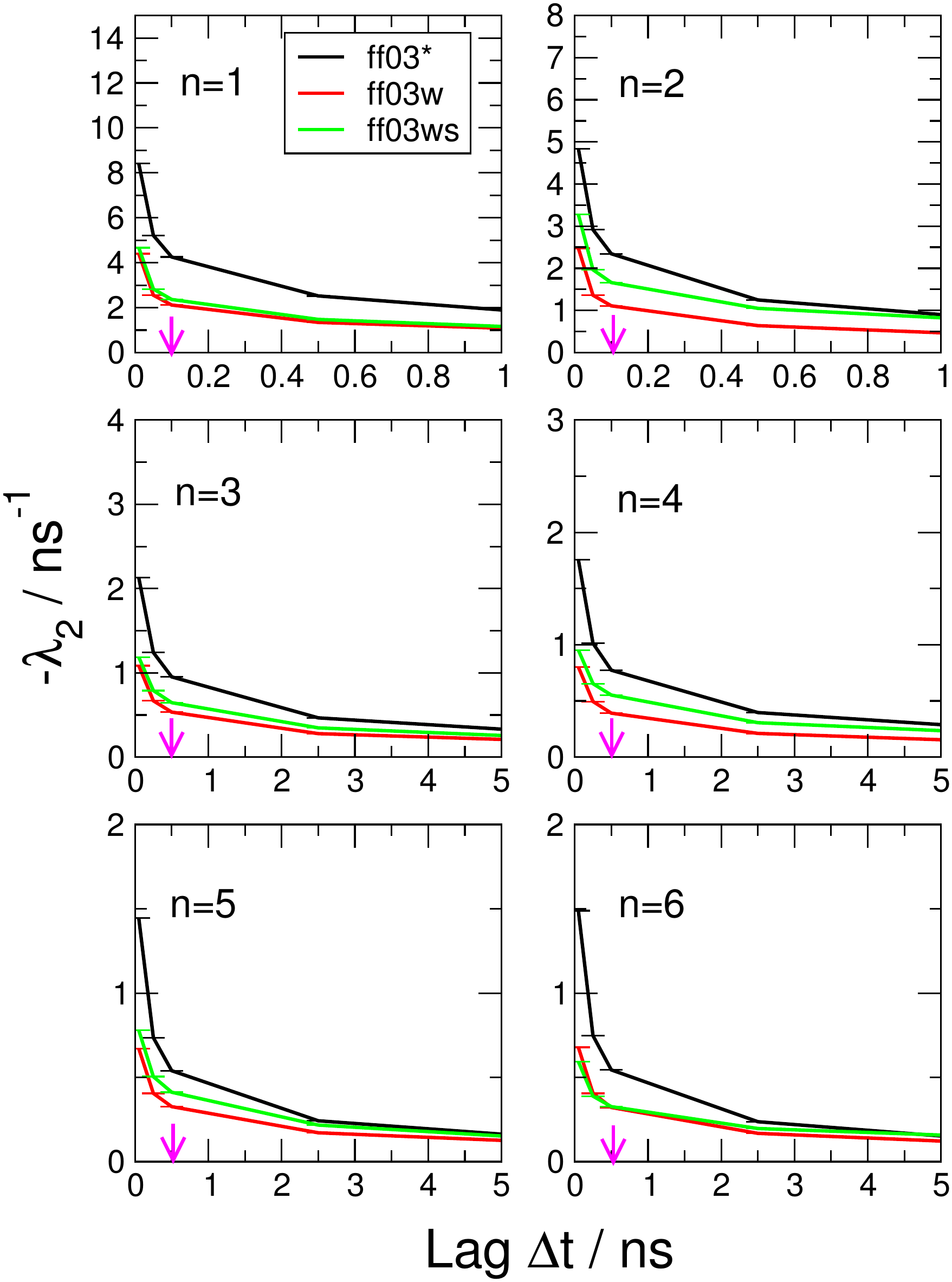}
\caption{Convergence of relaxation times with lag time $\Delta t$ for
the diffusion model.}
\label{fig:conv}
\end{figure}

Fitting of the 1D diffusion model followed a similar method to previous
work \cite{hummer-2005,best-2006-2,best-2011-3}. 
The all-atom simulation trajectories were projected onto the Trp-Cys
distance $r_\mathrm{cw}$, which was then discretized into regularly 
spaced bins spanning the populated range of that coordinate. The number of
bins used was 30 for $n=1-3$ and 60 for $n=4-6$. The increased number of
bins is required in order to described accurately the shorter distances
where contact formation occurs for the longer peptides. Count matrices
of the number of observations $N_{ij}$ from bin $j$ to bin $i$ after a 
given lag time $\Delta t$ were constructed. Best-fit diffusion coefficients 
and free energies were fitted to these data using the previously 
described Bayesian procedure. A lag time of $\Delta t=0.1$ ns was 
used for $n=1-2$ and a lag time $\Delta t=0.5$ ns for $n=3-6$, in
order that the resulting dynamics was approximately Markovian.
In Figure \ref{fig:conv} we show the convergence of the slowest 
relaxation time of the model with lag time. 
A smoothing prior on the similarity of diffusion coefficients $D_i$,
$D_{i+1}$ for adjacent bins $i$, $i+1$ of the form 
\begin{equation}
P(D_i,D_{i+1}) = \exp\Big[-\frac{(D_i-D_{i+1})^2}{2\gamma^2(\min(D_i,D_{i+1}))^2}
\Big]
\end{equation}
was used. A stiffness coefficient of $\gamma=0.1$ was chosen.

\begin{figure}[thbp]
\centering
\includegraphics[width=8 cm]{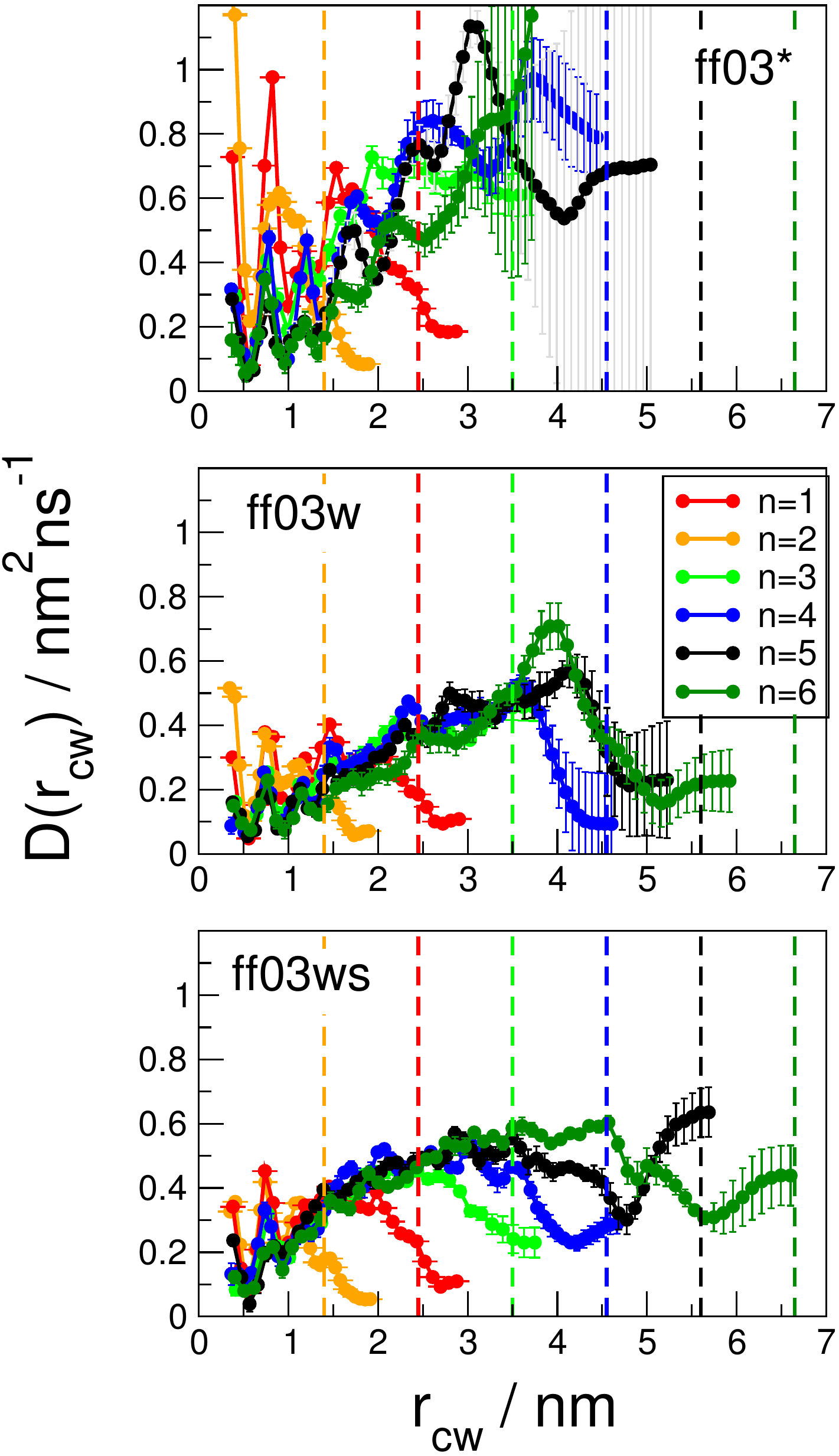}
\caption{Position-dependent diffusion coefficients from Bayesian
fit. Vertical broken lines give the approximate maximum extension for each
length $n_b$, given by $0.35~n_b$ nm.}
\label{fig:bayesdq}
\end{figure}

\section{Calculation of rates from diffusion model using SSS theory}

Given the one-dimensional $D(r_\mathrm{cw})$ and equilibrium distribution
$p_\mathrm{eq}(r_\mathrm{cw})$, the Szabo-Schulten-Schulten theory\cite{szabo-1980} allows 
the rate of diffusion-limited contact formation $k_{D+}$
to be calculated as \cite{lapidus-2002} (for instantaneous quenching
at $r_c$):
\begin{equation}
k_{D+}^{-1} = \int_{r_c}^{r_\mathrm{max}}
\frac{
\big[\int_r^{r_\mathrm{max}}p_\mathrm{eq}(s)ds\big]^2}
{D(r)p_\mathrm{eq}(r)}dr
\end{equation}
For a distance-dependent reaction rate $q(r_\mathrm{cw})$, the diffusion-limited
rate can be estimated via:
\begin{equation}
k_{D+}^{-1} = k_R^{-2}\int_{0}^{r_\mathrm{max}}
\frac{
\big[\int_r^{r_\mathrm{max}}(q(s)-k_R)p_\mathrm{eq}(s)ds\big]^2
}
{ D(r)p_\mathrm{eq}(r)) } dr
\end{equation}
where $k_R$ is the reaction-limited rate:
\begin{equation}
k_R = \int_0^{r_\mathrm{max}}p_\mathrm{eq}(r)q(r)dr
\end{equation}

The above integrals were evaluated over the discretized bins
of the diffusion model,
counting for the first bin only the range between $r_c$ and
the outer edge of the bin.

\section{Fitting of distance-dependent quenching rates to experimental data}

\begin{figure}[htbp]
\centering
\includegraphics[width=8 cm]{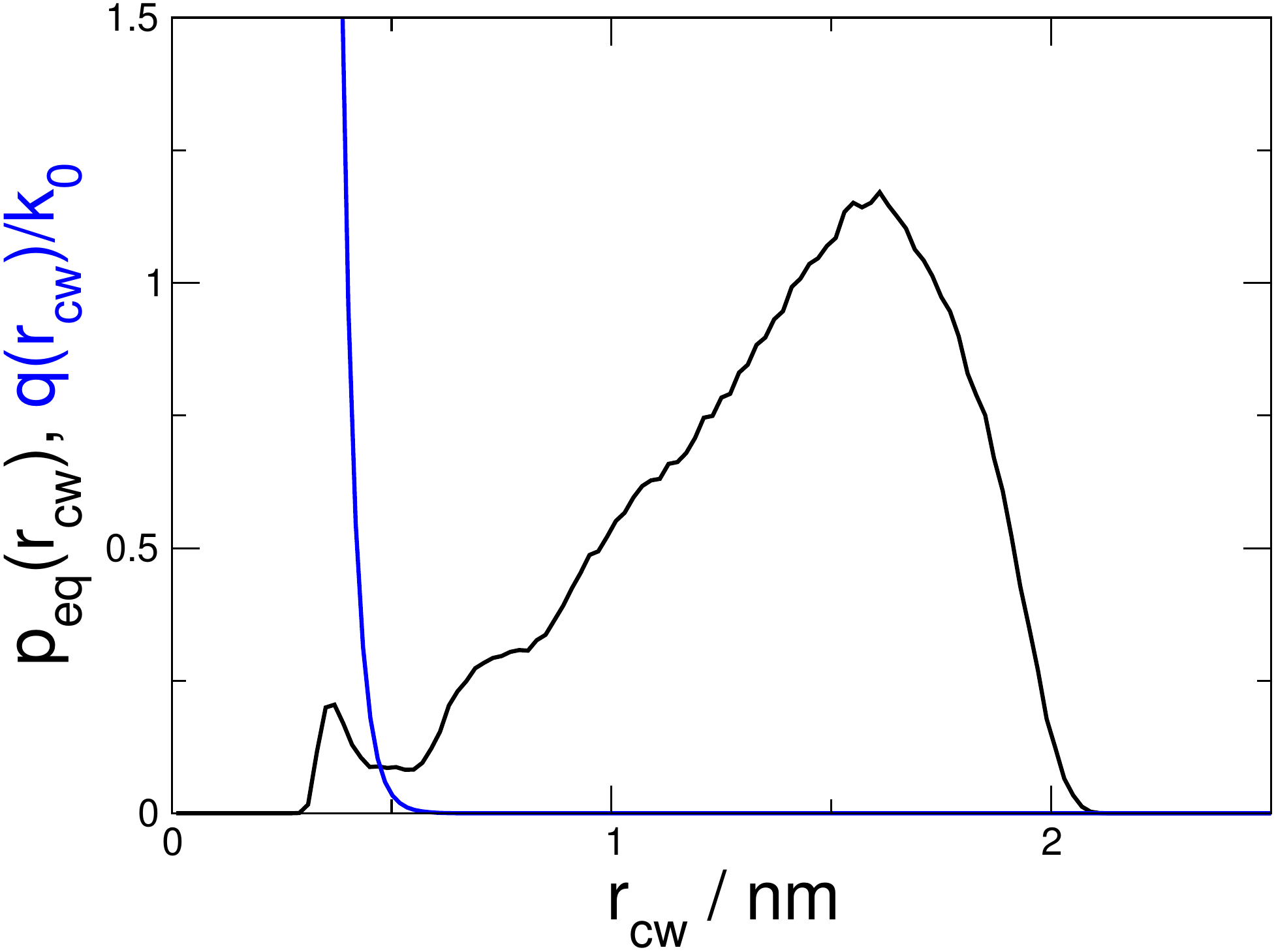}
\caption{Distribution function $p(r_\mathrm{cw})$ for the minimum distance
between N-acetyltryptophanamide (NATA) and zwitterionic cysteine from 
simulation in explicit water (black) and distance-dependence of
fitted quenching rate (blue).}
\label{fig:natacys_gr}
\end{figure}

Our initial model for the distance-dependent quenching rate was a step function
with parameters taken from the literature \cite{yeh-2002}. Since this is a 
fairly strong assumption about the distance dependence, we also tried
a more realistic alternative. Assuming that the 
quenching mechanism can be described as electron transfer, we modelled the 
distance-dependent rate $q(r_\mathrm{cw})$ with an exponential function:
\begin{equation}
q(r_\mathrm{cw}) = k_0 \exp[\beta(r_\mathrm{cw}-r_c)]
\end{equation}
We chose a contact distance of $r_c=0.4$ nm as before; 
the choice is quite arbitrary as an 
equivalent expression could be obtained with a different choice, by 
appropriately scaling $k_0$. We used the data reported by Lapidus \textit{et al.}
\cite{lapidus-2001}
to determine the optimal parameters. The main difference from their work is
that we use a pair distance distribution function determined from molecular 
simulation rather than a uniform distribution. We ran a 1 $\mu$s
of a single N-acetyl tryptophanamide (NATA) and zwitterionic cysteine in 
water. We did not attempt to include trehalose due to the requirement to
have a suitable force field for it, as well as the difficulty of sampling
such a system; we instead assume that the distribution in water is 
sufficiently similar. The probability density is shown in 
Fig. \ref{fig:natacys_gr}. Note that we have not attempted to 
normalize by the available volume at each distance as for a regular
pair distribution function, because that
operation is not easily defined for the minimum Trp-Cys distance we
use as a coordinate. Instead, we use observed density of distances $p(r_\mathrm{cw})$
directly; the drop-off at larger $r_\mathrm{cw}$ due to the finite
system size is unimportant since the quenching is so short ranged.

We fit the absorbance decays from Lapidus \textit{et al.}
\cite{lapidus-2001} to the function 
\begin{equation}
A(t)= A_0 L(t)\exp\Big[-\frac{Q}{Q_0}
\int_0^{r_\mathrm{max}}p_\mathrm{eq}(r)[1-S(t|r)]dr\Big]
\end{equation}
where $L(t)$ is the experimentally determined triplet lifetime in the
absence of Cys, $A_0$ is a scaling coefficient for fitting, $Q$ and 
$Q_0$ are the concentration in the experiment and the NATA-Cys 
simulation respectively, and $S(t|r) = \exp[-q(r)t].

\begin{figure}[htbp]
\centering
\includegraphics[width=8 cm]{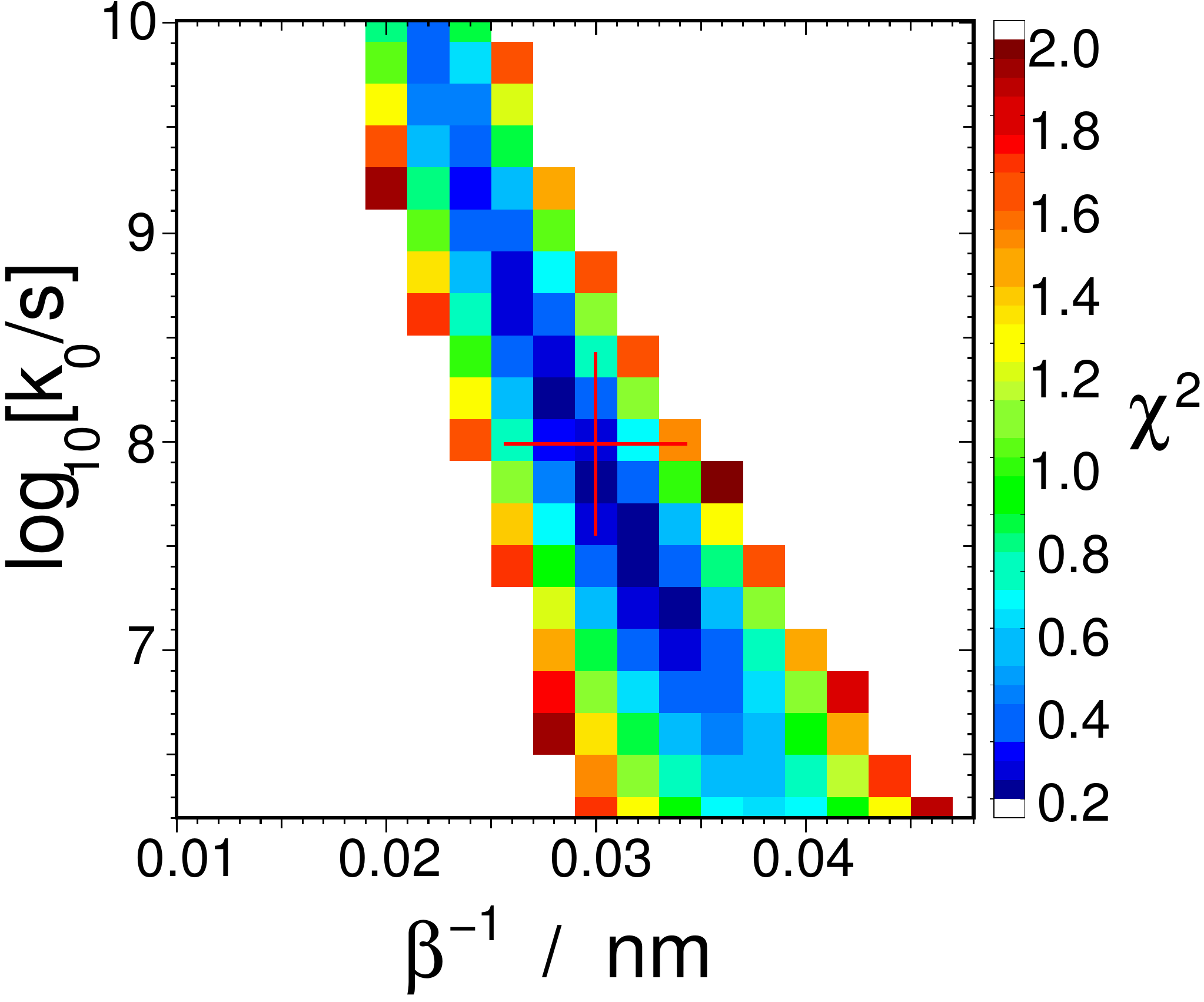}
\caption{$\chi^2$ surface for fit to Trp-Cys quenching data in a room
temperature trehalose glass. Red cross indicates chosen ``optimal''
parameters.}
\label{fig:chi2}
\end{figure}

\begin{figure}[thbp]
\centering
\includegraphics[width=8 cm]{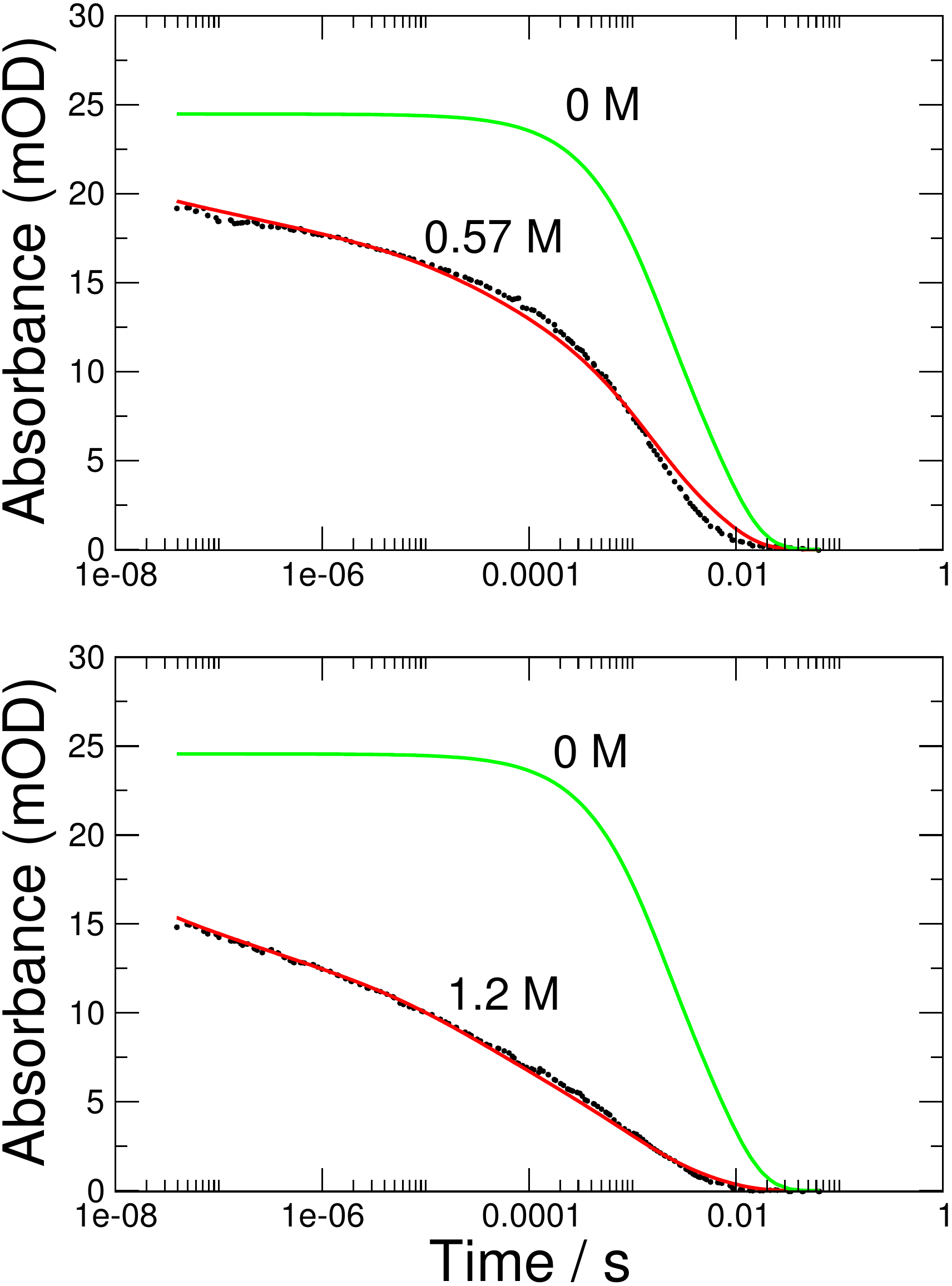}
\caption{Fit to raw data for Trp-Cys quenching from Lapidus 
\textit{et al.}\cite{lapidus-2001}. Green curve shows the survival $L(t)$
of the Tryptophan triplet in the absence of Cysteine, black curve is 
the experimental data in the presence of Cysteine concentration shown,
red curve is the fit with the optimal parameters for the distance
dependence of the quenching rate.}
\label{fig:lisadata}
\end{figure}

We systematically varied the two free parameters $\beta$ and $k_0$ on a grid,
obtaining an overall $\chi^2$ surface shown in Fig. \ref{fig:chi2}. 
While there is
clearly a lot of correlation between the parameters, we 
picked the values at the center of the optimal region as optimal values. Using
these values ($k_0=1.0 \times 10^8$ s$^{-1}$, $\beta=33.333$ nm$^{-1}$), we 
indeed find a very good fit to the experimental data, as shown directly
in Fig. \ref{fig:lisadata}. There is some discrepancy at long times for
the lower concentration, which has been explained in terms of
very slow diffusion within the glass \cite{lapidus-2001}, however this effect is less
important at the higher concentration. 

\clearpage

\bibliography{all}